# Two Trends of Composition Variation of Zircons and Their Significance in Origin Discrimination

Xuezhao Bao[1]

Department of Earth Sciences, the University of Western Ontario, 34-534, Platt's Lane, London, Canada, N6G 3A8

**Abstract:** Zircons can crystallize in a wide range of physical and chemical conditions. At the same time, they have high stability and durability. Therefore zircon can grow and survive in a variety of geological processes. In addition, the diffusivity of chemical compositions in their crystals is very low. Consequently, we can trace back the evolution history of the planetary materials containing zircon with zircon U-Th-Pb geochronology and geochemistry studies. However, this depends on our ability to decipher its genesis, namely magmatic or metamorphic origins. In this paper, we have found that there are obvious differences between magmatic and metamorphic zircons in their chemical composition zonations. The magmatogenic/magmatic zircons exhibit composition zonation of increasing $HfO_2$, and ($UO_2 + ThO_2$) content and decreasing $ZrO_2/HfO_2$ ratio and $ZrO_2$ content from inner to outer parts within each growth zone or from core to rim of a crystal. The metamorphogenic/metamorphic zircons exhibit compositional variation trend opposite to that of magmatic (igneous) zircons, tending to decrease in $HfO_2$, ($UO_2+ThO_2$) and increase in $ZrO_2/HfO_2$ ratio and $ZrO_2$ from core to rim of a crystal. These chemical composition variation trends are thought to be controlled by the crystal-chemical features of ions themselves and the evolution trends of magmatism and metamorphism respectively, and can be used to identify the genesis of zircons. Their morphological features are also discussed.

We have found two kinds of composition zonations in zircons through electron microprobe analyses. These composition zonations may reflect the geological evolution history of their parent rocks and can be used to identify the origin of zircons.

## 1. GEOLOGICAL OCCURRENCE OF ZIRCONS AND ANALYSIS WORKS

The zircons used in this paper are listed in table 1. All metamorphic zircons come from metamorphic rocks. Their origins have been determined by zircon morphology and geological field settings combined with their single zircon U-Th-Pb isotope ages, which fit the metamorphic events determined by field observations. In addition, zircons 87013 and 937 have been concluded to be of metamorphic origin by Wang [1] and Gan [2] respectively through zircon morphology, U-Th-Pb isotope ages and field observations. Zircons 084, 2155 and G2126 possess rounded cores (2 in fig. 1) or rounded zonings and cores (3 in fig.1). These rounded cores may be the remnants of old zircons or crystallized in an earlier stage, and then were corroded into rounded crystals by the metamorphic fluid. New zircon crystals can continue to grow around these cores. However, metamorphic fluid with increasing temperature can corrode these new growth parts into rounded shapes again, and then newer parts can grow around these older parts. With these repeated growths, a zircon crystal with a rounded core and zonings can be formed. These crystals also have accumulative crystallization and partially rounded shapes (4 and 5 in fig.1). These can indicate their metamorphic or metamorphic re-crystallization origin. All of these magmatic zircons exhibit euhedral growth zonings and crystal shapes, which can indicate their magmatic or igneous origin.

**Table 1. The geological occurrence, U-Th-Pb age, and genetic features of zircons***

| genesis | Sample # | location | Rock type | U-Th-Pb age/Ma** | Morphology |
|---|---|---|---|---|---|
| Metamorphic | 87013 | Danzhu, Longquan, Zhejiang | medium-grade metamorphic granodiorite | 1878 ± | ref(1) |
| | 937 | Yeiken, Jianou, Fujian | low-medium-grade metamorphic quartz porphyry | 773 ± | ref (2) |
| | 084 | Xinhe, Inner Mongolia | medium-high-grade sillimanite garnet gneiss | 1900 ± | multi- rounded cores and glomerocrysts (2 in fig.1) |
| | 2155 | Xinhe, Inner Mongolia | high grade granulite | 1900 ± | multi- rounded cores and glomerocrysts (4&5 in fig.1) |
| | G2126 | Xinhe, Inner Mongolia | high grade granulite | 1900 ± | rounded shape and multi-rounded cores (3 in fig.1) |
| magmatic | H903 | Mashan, Hunan | granite | 216± | euhedral magmatic growth zonings |
| | 9303 | Xiao Qinling | pegmatite | 1850± | long columnar, euhedral shape |
| | H9307 | Xiao Qinling | pegmatite | | long columnar, euhedral shape |
| | H006 | Banxi Group, Hunan | andesitic agglomerate | 933± | euhedral magmatic growth zonings |
| | T9305 | Langfang, Hebei | plagioclase amphibole gneiss | | euhedral magmatic growth zonings(1 in fig.1) |

**\* The U-Th-Pb zircon ages are determined by Huiming Li in the Geochronology and Isotope Geochemistry lab, Tianjin Institute of Geology and Mineral Resources.**
**\*\* Ma = million years**

These samples were mounted in epoxy, and then grinded and polished for electron microprobe (EMP) analyses. We used a JEO EMP in the Chinese University of Geology in Beijing and another JEO EMP in Tianjin institute of Geology to do the analysis. Seventeen elements were analyzed. Standard materials were also analyzed for each element to correct and calculate the concentration of these elements. The main standards we have used are: zircon for Zr, 100% metal Hf for Hf, 100 metal U for U, and 87.88% $ThO_2$ for Th. Data from 187 analyzed points were collected. The analysis conditions are the same in both machines: analyzed voltage: 15 KV, current: $2 \times 10^{-8}$ A, analyzed speed: 100 seconds /point. The ZAF calibration method was used and the errors are within the range of $400 \times 10^{-6} \pm wt\%$.

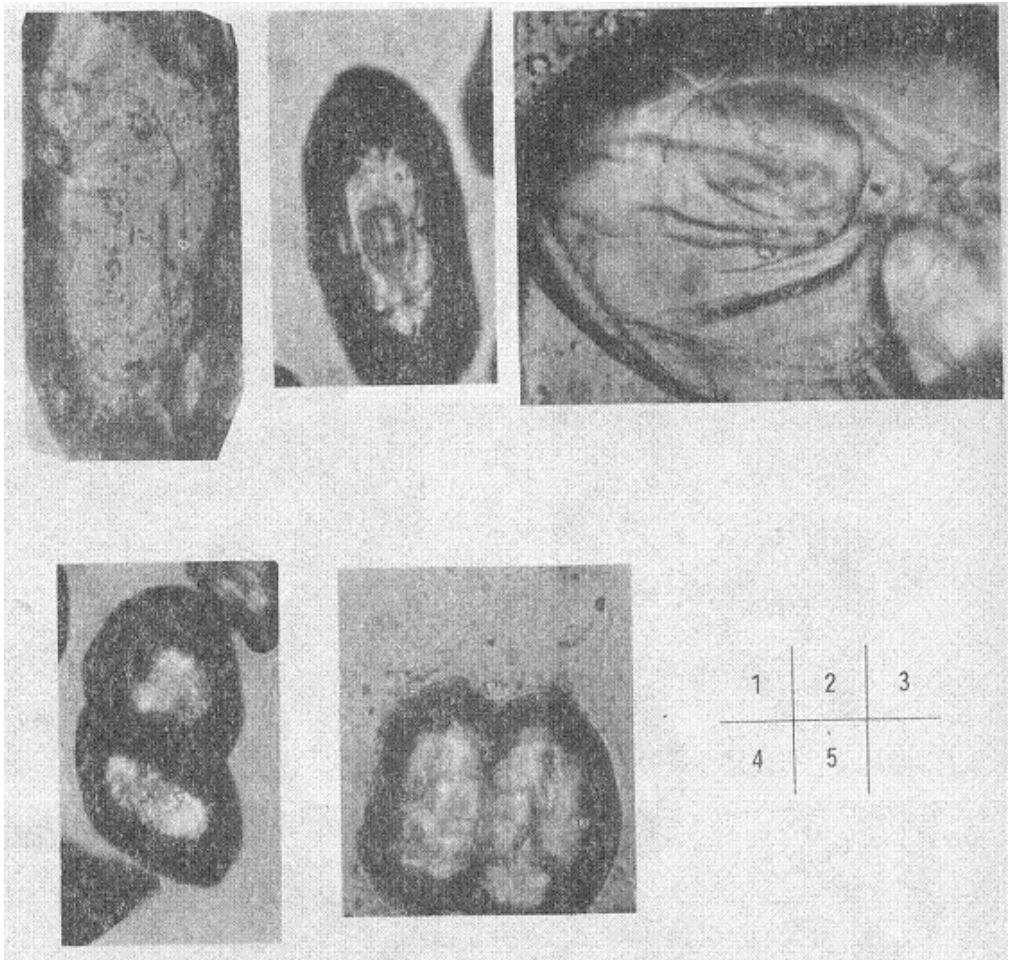

**Fig. 1. The microscopic photos of zircons**
1. Zircon T9305 with euhedral growth zonings. Magnification X1000. 2. Zircon 084 with rounded core and growth zonings. Magnification X400. 3. Zircon G2126 with rounded core and growth zonings. Magnification X1000. 4 and 5: Zircons 2155 with rounded shapes and glomerocrysts. Magnification X400.

## 2 RESULTS AND DISCUSSION

2.1 magmatic zircons

The magmatic zircons exhibit a composition zonation trend of increasing $HfO_2$ and $(UO_2+ThO_2)$ concentrations, but decreasing ratio of $ZrO_2/HfO_2$ from core to rim of the crystals without obvious growth zonings, or from the inner to outer parts within each growth zoning of crystals as illustrated in figs. 2, 3 and 4. This is consistent with the composition analysis results described in several previous studies on zircons from granitic rocks. Wang et al. [4,5] reported that the zircons from a granitic rock originating from the crust and another granitic rock originating from the mantle exhibit a composition zonation trend of decreasing ZrO$_2$/HfO$_2$ ratio, but increasing HfO$_2$ concentration from the core to the rim of a crystal. Zircons with growth zonings also exhibit a composition zonation trend of decreasing ZrO$_2$ concentration, but increasing UO$_2$ and HfO$_2$ concentrations from inner to outer parts in each zoning. Zhou also reported that a group of magmatic zircons with a high content of Hf have a concentration of HfO$_2$ ~3 weight (wt) % in their centers, but ~7 wt% in their rims. He concluded that this trend is produced by Zr entering the zircon crystals relatively earlier than Hf in the process of magma crystallization and differentiation [7]. Pagel reported that all zircons (no matter how large or small in size from a granite sample) exhibit a UO$_2$ concentration of 0.03wt%~0.5wt% in their cores, but ~3.5wt% in their rims. The HfO$_2$ concentration is higher in their rims [8]. Bibikova concluded that the zircons from all granitic rocks possesses a composition zonation trend of increasing U and Th concentrations from core to rim of a crystal [9]. Compared with the chemical composition evolution trends of different magmatic rocks and their zircons, the following two additional points should be noted:

(1) A decreasing trend of the ZrO$_2$/HfO$_2$ ratio from the core to the rim of a magmatic zircon crystal described above is consistent with a decreasing trend of the ZrO$_2$/HfO$_2$ ratio of zircons from mafic, intermediate to felsic rocks [10].

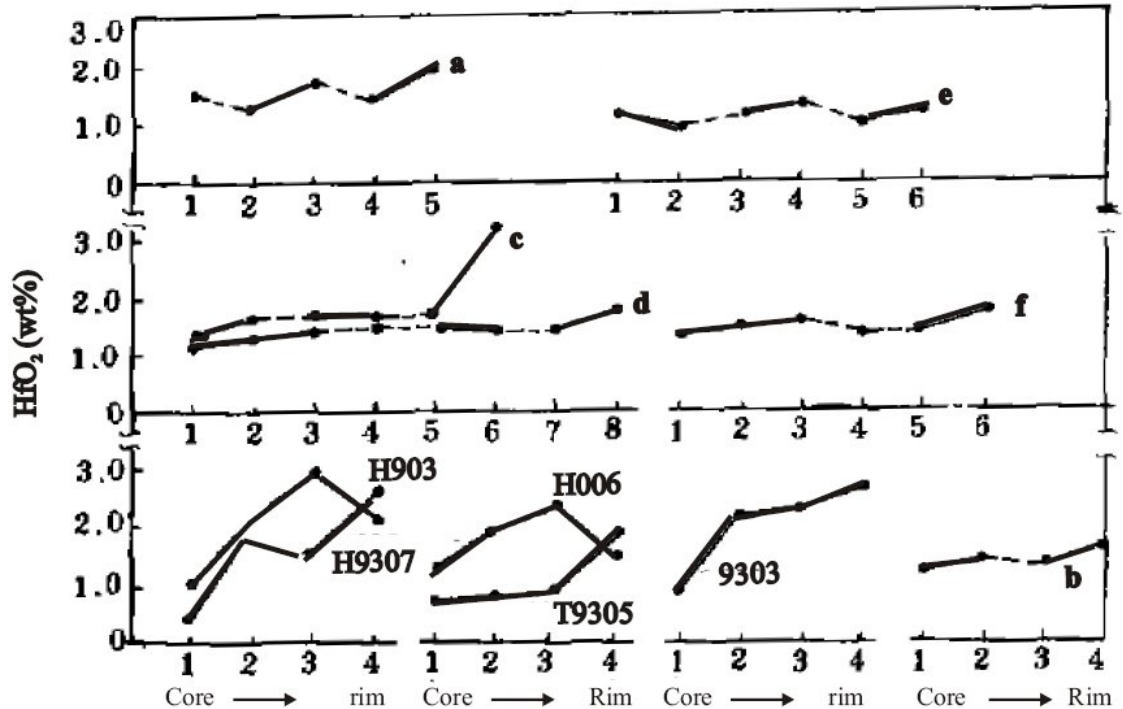

**Fig.2. the content variations of HfO$_2$ from core to rim of magmatic zircon crystals. The solid lines signify the growth zonings in zircons a ~ f. Data of a~f from ref. 6. Same in figs 3 and 4.**

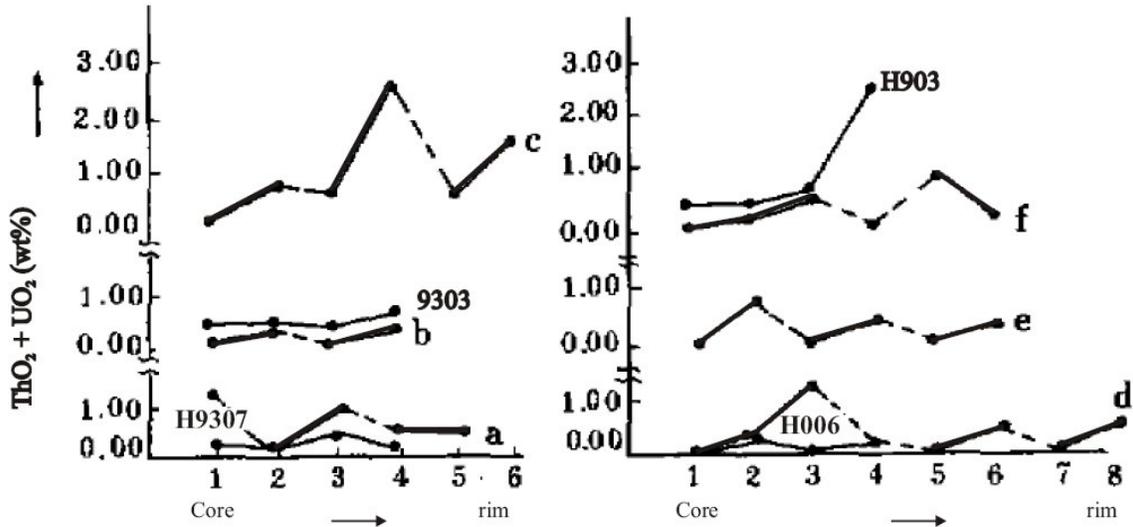

Fig. 3. The content variations of ($UO_2$ + $ThO_2$) from core to rim of magmatic zircon crystals.

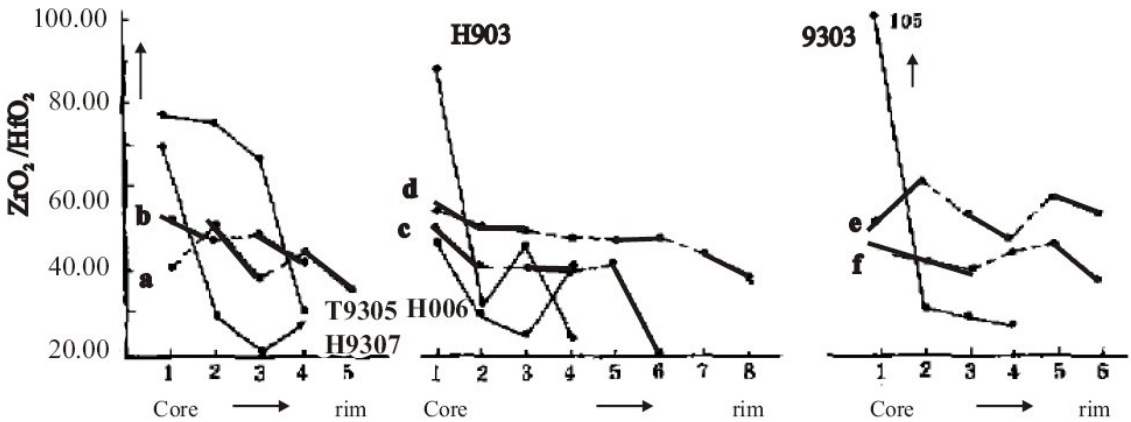

Fig. 4. The $ZrO_2/HfO_2$ ratio variations from core to rim of magmatic zircon crystals.

(2) An increasing trend of $UO_2$ + $ThO_2$ content from core to rim of a magmatic zircon crystal described above is consistent with an increasing trend of the $UO_2$ and $ThO_2$ contents from ultra-mafic, mafic, intermediate to felsic rocks[11].

This indicates that the composition zonation trends of these zircons reflect the evolution trends of the magmas they crystallized in, since magmas usually evolve from relatively mafic to relatively felsic compositions in crystallization and differentiation processes. At the same time, the composition zonations of a zircon crystal also depend on its ionic properties. According to Goldschmidt's rule, $Zr^{4+}$ with a small radius (R) of 0.098 nm enters zircon crystals relatively earlier than the large ions $U^{4+}$ (R = 0.114 nm) and $Th^{4+}$ (R = 0.118 nm). $Zr^{4+}$ with a small electronegativity of 1.21 enters zircon crystal a little earlier than $Hf^{4+}$, with a relatively large electronegativity of 1.38 according to Ringwood's modification to Goldschmidt's rule. This produces a relatively high $Zr^{4+}$ content in the core (the earlier crystallized part) but a relatively high content of $U^{4+}$, $Th^{4+}$ and $Hf^{4+}$ in the rim (the later crystallized part) of a magmatic zircon crystal. The composition evolution of magmatic rocks and the crystal chemical properties of Zircons, such as Zr, Hf, U and Th, are relatively stable. Therefore, the composition zonation trends, especially the $HfO_2$, ($UO_2+ThO_2$), and $ZrO_2/HfO_2$, of zircons are generally fixed, and consequently can be used as the indicators for magmatic (magmatogenic) zircons.

As to the growth zonings of zircons a~f, I think they may have undergone multistage growths induced by multistage magma activities during the magma crystallization and differentiation process(es); since the composition variation trend in each growth zoning is consistent with the composition zonation trend of magmatic zircons as shown in figs. 2, 3 and 4, and the composition variation trends of zircons from mafic, intermediate to felsic rocks as discussed above. In addition, the two growth zonings in the most inner parts of zircon e exhibit mainly (311) pyramid faces, indicating that the two most inner zonings crystallized in magmas with relatively high mafic compositions according to crystal morphology [6]. On the contrary, in the outer growth zonings, the (311) pyramid faces disappear and are replaced by (111) pyramid faces, indicating that these outer growth zonings crystallized in magmas with relatively high felsic compositions according to crystal morphology [6]. Therefore, from inner to outer zonings, the magmas they crystallized in are different. Namely, these growth zonings crystallized in different magmatic activity stages.

The composition variations in the edges of zircons H9307, H006, a and e are not consistent with the major parts of crystals. First, these abnormal composition variations may be produced by later geological activities, since all of these occur in the edges of crystals. For instance, only the two analyzed points in a and e zircons have composition variation abnormalities among six analyzed zircon crystals, and only these two analyzed points are close to the edges of crystals as indicated in the figures of ref. 6. The composition variation trends of the growth zonings inside the crystals in the six samples are normal as indicated in figs. 2, 3 and 4.

The composition variation abnormalities also only occur in the edges of zircons H9307 and H006 as shown in figs. 2 and 3. H9307 are the zircons coming from a pegmatite. After the crystallization and differentiation of the pegmatic magma, these zircon crystals may have undergone re-crystallization in their edges from the high temperature hydrothermal solutions in later stages. H006 are the zircons from a volcanic extrusive rock. After crystallizing from the volcanic extrusive rock, the edges of zircons may have undergone chemical alternation between crystals, water and fluid from the surrounding rocks and the Earth's surface. Zircons e is from a remelting/anatexis granite [6]. The $HfO_2$ and $ZrO_2/HfO_2$ variation abnormalities in the most inner zoning may have come from an early anatexis in the time when this zoning crystallized.

2.2 Metamorphic zircons

All of these metamorphic zircons exhibit composition zonations of increasing $ZrO_2$ concentration and the ratio of $ZrO_2/HfO_2$, decreasing $HfO_2$ and $UO_2 + ThO_2$ concentrations from core to rim of a crystal as indicated in figs. 5 and 6, which is opposite to those of magmatic zircons. Therefore, we can conclude that the crystallization conditions for metamorphic zircons are very different from those of magmatic zircons. Magma crystallization and differentiation is a process of decreasing temperature from liquid magmas to solid rocks. However, metamorphism is a process of mainly increasing temperature. With an increase of metamorphic temperature or grade, re-crystallization will occur, and metamorphic fluids will be produced in the parent rock. Further, selective fusion will occur and a magma-like fluid will form. From here it can be seen that there is not only a trend of increasing temperature, but also change from the solid parent rock to liquid magma during the process of metamorphism. This trend is opposite to the magmatism of changing from liquid magma to solid rock. Therefore, two opposite composition zonation trends can be observed in the zircons crystallized in the two processes.

2.2.1 The influence of composition variation of rocks on the composition variation of their zircons

After investigating metamorphic rocks with different metamorphic grades, it is concluded that with an increase of metamorphic grade, the Zr concentrations of rocks increase, and all of them enter the zircon crystals. Until the temperature reaches the hypersthene grade, the Zr concentration appears to be decreasing as illustrated in fig. 6 [12]. Therefore, zircons 87013 and 937 from rocks with medium and low-medium metamorphic grades respectively exhibit only an increasing trend of $ZrO_2$ from core to rim of a crystal. However, the zircons 2155 and G2126-1 (the rim parts of zircon G2126-3 have not been analyzed because of its poor crystal surface) from granulites with high metamorphic grades exhibit increasing trends in their inner parts, but decreasing trends in their outer parts. This is consistent with the $ZrO_2$ content of metamorphic rocks in an order of increasing metamorphic grade: namely $ZrO_2$ concentrations increase first and then decrease slightly as illustrated in figs. 6 and 7. In addition, a previous study also confirmed that with an increase in metamorphic grade, the U concentration of metamorphic rocks decreases [13]. Another study also found that the U and Th concentrations of a metamorphic quartz-mica diorite decrease after its metamorphism rising from the amphibolite facies to the granulite facies [14]. The generally decreasing trends of $UO_2 + ThO_2$ from the core to the edge of metamorphic zircon crystals reflect these composition variation trends. In short, the composition variation trends of metamorphic zircons reveal the composition variation (such as Zr, U and Th) of their parent rocks.

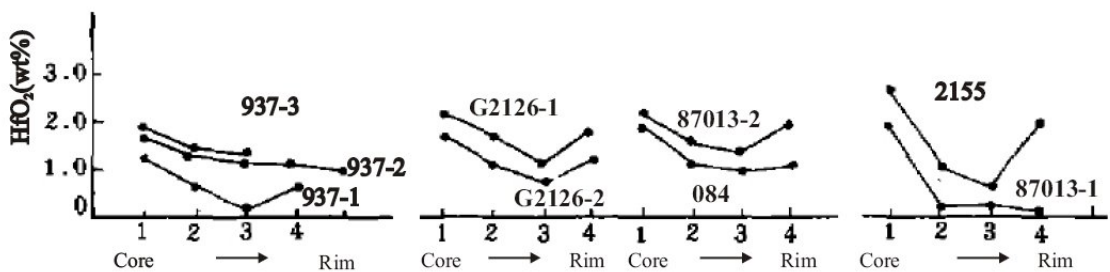

**Fig. 5. The content variations of $HfO_2$ from core to rim of metamorphic zircons**

2.2.2 Causes of compostion zonation abnormalities in metamorphic zircons

The composition zonation trends in the rim/edge of metamorphic zircons, such as G2126-1, 2155, 87013, are not consistent with their general composition zonation trends. The following may be the causes:

(1) When the rim parts of these metamorphic zircons began to crystallize, their parent rocks had mostly undergone metamorphism from medium to high grades. As discussed above, the Zr concentration of their parent rocks have begun to decrease as indicated in fig. 7, which will lead to a decrease of the Zr concentration of zircons crystallized in these rocks. Because $Hf^{4+}$, $U^{4+}$ and $Th^{4+}$ have an isomorphous replacement relationship with $Zr^{4+}$ in zircons, a variation of Zr would certainly causes a variation of Hf, U and Th composition in a zircon crystal.

(2) Under retrograde metamorphism conditions, composition zonation abnormalities can occur in the rim of garnets from metamorphic rocks [15]. The granulites, parent rocks of zircons G2126 and 2155 have obvious indicators of retrograde metamorphism. This may have induced the composition zonation abnormalities in their edges.

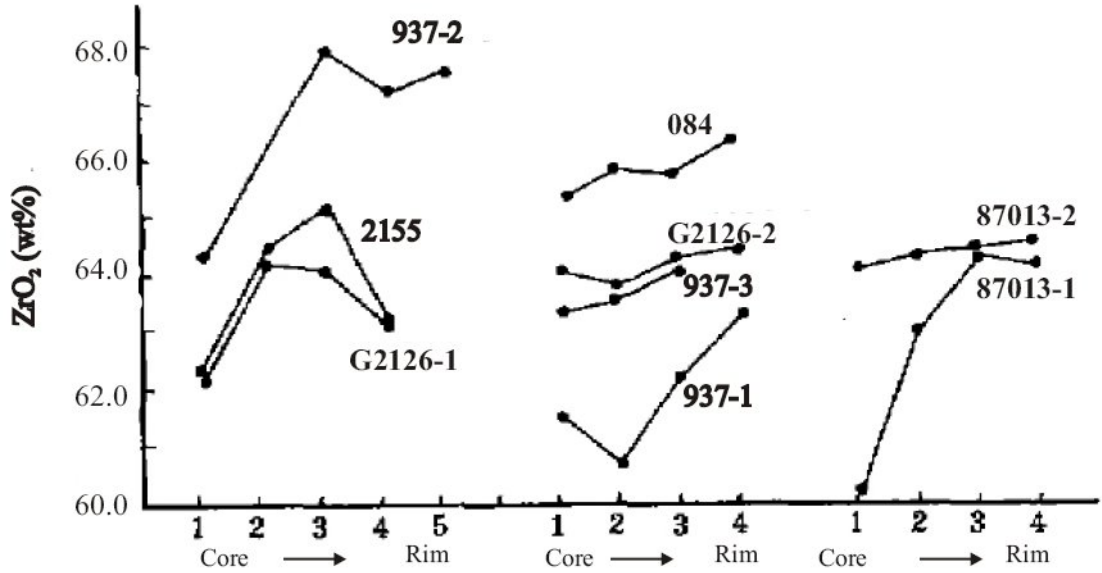

**Fig. 6. The content variations of $ZrO_2$ from core to rim of metamorphic zircon crystals**

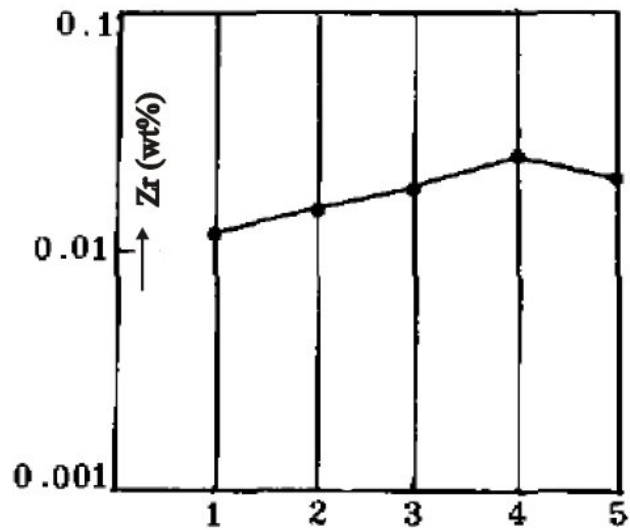

**Fig. 7. The content variations of Zr from low-grade (1) to high-grade (5) metamorphic rocks. Data from ref. 12.**
1. garnet zone ; 2 staurolite-andalusite zone ; 3. sillimanite-dolomite zone; 4. sillimanite-K-feldspar zone; 5. hypersthene zone.

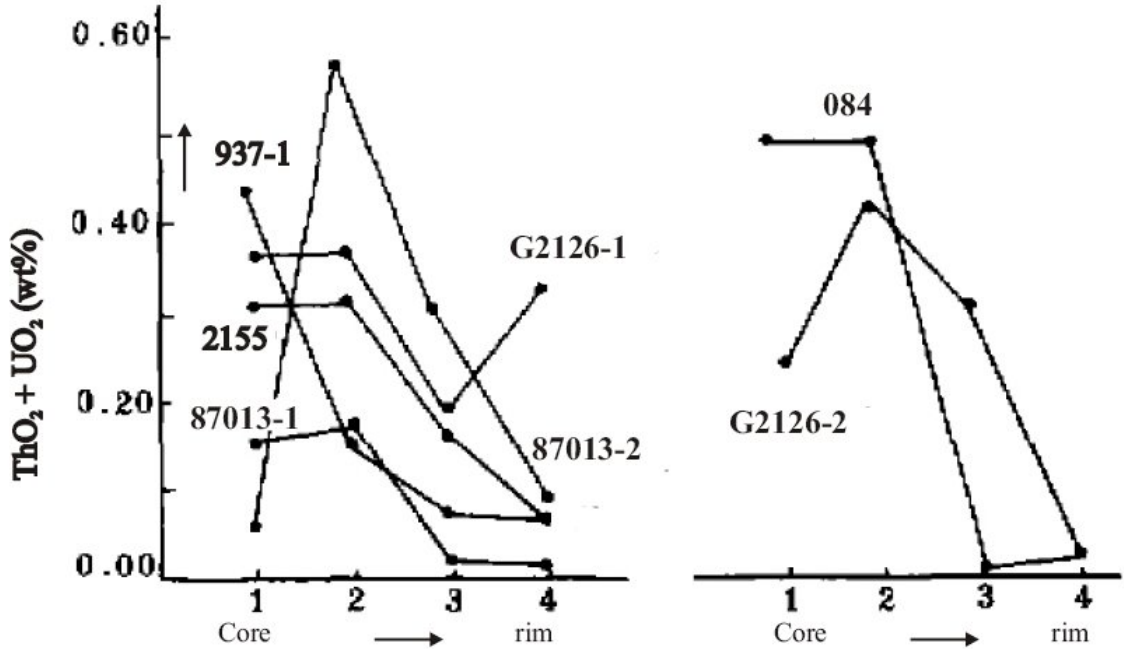

**Fig. 8. The ($UO_2$ + $ThO_2$) content variations from core to rim of metamorphic zircons.**

(3) The rocks containing these zircon samples were collected from an area close to the surface of the Earth. As with the magmatic zircons discussed before, the composition zonation abnormalities in the edges of these metamorphic zircons may be produced by weathering, rainwater erosion and etc. after their crystallization.

(4) Metamorphic re-crystallization zircons usually re-crystallize from former remnant zircons. Therefore, their cores may be remnants of the older zircons. This can produce composition zonation abnormalities in their centre, such as zircons 937-1 and G2126-2.

(5) After reaching granulite facies, the metamorphic temperature is relatively high. In the peak metamorphic stage, there is a trend of changing from metamorphism to magmatism. In this case, the composition zonation in the edges of metamorphic zircons may change from "metamorphic zircon type" to "magmatic zircon type".

In a word, the composition zonation abnormalities mostly occur at the edges of zircons, and can not have an influence over their general composition zonation trends.

## 3. CONCLUSIONS

(1) Euhedral growth zoning is one of the indicators of magmatic zircons, but a rounded core and rounded growth zoning is one of the indicators of metamorphic zircons.

(2) In chemical composition zonations, the magmatogenic zircons exhibit composition zonation of increasing $HfO_2$ and $UO_2+ThO_2$ contents, and decreasing $ZrO_2/HfO_2$ ratio and $ZrO_2$ content from the inner to outer parts of each growth zoning or from core to rim of a crystal. The metamorphic zircons exhibit a compositional variation trend opposite to that of magmatogenic zircons, which generally tends to decrease in $HfO_2$, ($UO_2$+ $ThO_2$), and increase in $ZrO_2/HfO_2$ ratio and $ZrO_2$ from core to the edge of a crystal.

**<u>Notes:</u>**

This is my first paper on zircons, which was originally published in Acta Mineralogica Sinica, 15 (4): 404-410 (1995).

These composition zonations have been reexamined by electron microprobe and Raman microprobe analyses in:

1. Bao and Gan, The Minerageny of Two Groups of Zircons from Plagioclase-Amphibolite of Mayuan Group in Northern Fujian, Acta Petrologica Et Mineralogica, 1996, Vol. 15, No. 1, 73-79, or at http://arxiv.org, arXiv: 0707.3181, July 2007.

2. Bao et al., A Raman spectroscopic study of zircons on micro-scale and its significance in explaining the origin of zircons, Scientia Geologica Sinica, 1998, Vol.33 (4): 455-462, or at http://arxiv.org, arXiv: 0707.3184, July 2007.

In addition, a decreasing trend in crystallization temperature from core to rim of magmatic zircon crystals described in our papers is consistent with the calculated core-to-rim variation in crystallization temperature of a 4.3 billion year old magmatic zircon (Watson and Harrison, Science, 2005, V308, 841-844).

It is widely accepted that zircons can record the evolution history of the Earth and planets. For instance, through the studies of geochronology and isotope geochemistry of 4.0 - 4.4 billion years old zircons, it was derived that continental crusts began to form in Earth's first 100 million years (Ma) (Watson and Harrison, same as above; Harrison et al., Science, 2005, V310, 1947 – 1950), although they may have developed mainly around 2700 Ma and 1900 Ma ago (Condie et al., Precambrian Research, 2005, V139, 42-100).

The U, Th zonations of zircons described in our papers actually reflect the migration trends of U and Th in the Earth. U and Th are the most important heat-producing elements in the Earth and other terrestrial planets, and are the main energy sources for planetary evolution. At the same time, the X-rays released from them may have an important influence on life beings. We have proposed theoretical migration models and distribution patterns of U and Th in Earth and other terrestrial planets. Their influence on the Earth and terrestrial planetary dynamics, and the origin and evolution of life (including a graphic summary of zircon composition zonations and morphologies) has also been discussed in:

X. Bao & A. Zhang, Geochemistry of U and Th and its influence on the origin and evolution of the Earth's crust and the biological evolution. Acta Petrologica et Mineralogica, 17(2): 160-172 (1998), or http://arxiv.org, arXiv: 0706.1089, June 2007.

**Acknowledgements**

I would like to thank Dr. RA Secco for offering me a postdoctoral position, which provoked my interest in translating this paper in my after-work time. This original work was supported by a grant from the National Natural Science Foundation of China (49202021).